\documentclass[preprint,aps,showpacs]{revtex4}
\usepackage{graphicx,amsfonts}
\usepackage{epsfig}

\begin{document}

\title{Evading the Few TeV Perturbative Limit in 3-3-1 Models}

\author{Alex Gomes Dias\footnote{e-mail: alexdias@fma.if.usp.br}}
\affiliation{Instituto de F\'\i sica, Universidade de S\~ao Paulo, \\
C. P. 66.318, 05315-970 S\~ao Paulo, SP, Brazil. }

\date{\today}
\begin{abstract}
Some versions of the electroweak SU(3)$_L\otimes$U(1)$_X$ models cannot be treated 
within perturbation theory at energies of few TeV. An extended version for these 
models is proposed which is perturbative even at TeV scale posing no threatening 
inconsistency for test at future colliders. The extension presented here needs 
the addition of three octets of vector leptons, which leave three new leptonic 
isotriplets in the SU(2)$_L\otimes$U(1)$_Y$ subgroup. 
With this representation content the running of the electroweak mixing angle, 
$\theta_W (\mu)$, is such that $\sin^2\theta_W(\mu)$ decreases with the increase 
of the energy scale $\mu$, when only the light states of the Standard Model 
group are considered. The neutral exotic gauge boson $Z^\prime$ marks then 
a new symmetry frontier.

\end{abstract}
\maketitle

\section{Introduction}
\label{sec:intro}

The known 3-3-1 models proposed nearly twelve years ago are good 
candidates to describe some new phenomena we expect to see in the near 
future \cite{331}. They offer a novel phenomenology like, for example, being an 
appropriate framework to study in details bilepton through $e^-e^-$ production 
in linear colliders \cite{rasin}, also within other observational capabilities 
\cite{dion,maisbi}. They provide a natural theoretical explanation to the 
families replication problem. 

Despite these and other curious features the early model versions present a severe 
limitation. The problem is that, like in any Standard Model gauge group extension, 
they predict a bunch of new particles. If from one side we have a base for 
computing new states production, on the other side we must worry about the consequences 
of the additional degrees of freedom. New particles already arising at the TeV scale 
affect significantly the coupling evolution with energy. In fact, even if 
these states do not appear directly as on shell states, their degrees of freedom 
effectively contribute to the dynamics according to the renormalization group 
equations. This problem is particularly dramatic in the model versions we shall 
discuss here. Such versions belong to the class of 3-3-1 models where the third 
component in the lepton triplet is a positive charged particle \cite{331,pt}. 
There the SU(3)$_L\otimes$U(1)$_X$ gauge coupling constants, denoted 
by $\alpha_L$ and $\alpha_X$ respectively, are linked by the following relation 
involving the electroweak mixing angle $\theta_W$
\begin{equation}
\frac{\alpha_X(\mu)}{\alpha_L(\mu)}=
\frac{\sin^2\theta_W (\mu)}{1-4 \sin^2\theta_W(\mu)}.
\label{polo}
\end{equation}
\noindent
This relation is valid above the energy scale $\mu\geq\mu_{331}$ 
where all symmetries become evident inside this theoretical framework. Evidently, the 
symmetry breakdown SU(3)$_L\otimes$U(1)$_X\rightarrow$ SU(2)$_L\otimes$U(1)$_Y$ has 
to be postulated to have occurred when a scalar field $\chi$ condensates with 
$\langle\chi\rangle\simeq \mu_{331}$. But Eq. (\ref{polo}) requires 
$\sin^2\theta_W <1/4$, which is in accordance with direct measurement.  
Eq. (\ref{polo}) when confronted with the current value 
$\sin^2\theta_W(M_Z)=0.23113 (15)$ \cite{pdg} gives rise to a 
trouble in these models. The point is that their SU(2)$_L\otimes$U(1)$_Y$ effective 
theory has almost the same Standard Model minimal particle content, apart from few 
additional scalar multiplets. Therefore, the renormalization group equation 
solutions in this case show us that $\sin^2\theta_W$ increases with 
energy. It means that Eq. (\ref{polo}) points to a nonperturbative regime at energy 
values which we shall be just interested. The equation defines the initial value 
$\alpha_X(\mu_{331})$ and since it comes from an abelian group it also increases 
with energy, being close to one for energies around few TeV in the minimal models. 
Consequently it is a serious problem for the perturbative approach and it could make 
the models less appealing.

For example, running  $\sin^2\theta_{_W}$ with only the minimal Standard Model 
representation content we see that it reaches the value $0.25$ at the energy 
scale $\mu\sim 4$ TeV \cite{phf98}. A study of perturbative limits in 3-3-1 models 
was carried out in \cite{pl331}. There it was shown that perturbation theory cannot 
be used already above scales of few TeV if we consider the minimal versions with or 
without supersymmetry.

There are some 3-3-1 constructions which do not have the problem above  
\cite{outros331}. But the ones considered here account for doubly charged bileptons 
intermediating processes where partial lepton number is violated, representing a very 
distinct signal to search for \cite{dion,rasin}. Other aspects and recent work on 
these models can be found in Ref. \cite{outros}.

In this work it is proposed an additional matter content to the early 3-3-1 models 
where perturbative treatment can be applied without such tight restrictions. 
We see that the simplest representations to be added to the minimal versions are 
four octets, three of vectorial leptons plus and a scalar one. With  them there is a 
prediction of a decreasing behavior for $\sin^2\theta_{_W}$ in a certain range of energy. 
The new exotic leptons are interesting also from the phenomenological point of view 
under the Standard Model gauge group in the see-saw mechanism \cite{foot}, and also 
under the extended group we consider here \cite{seesaw331}. 

The work is organized as follows. In Sec. \ref{sec:sec2} we study the renormalization 
group equations to understand how the nonperturbative limit appears in the minimal 
models. We then define a condition which should be satisfied in the model 
extensions, in order to make them perturbatively safe. We, then, justify the choice  
of the new representations content to be introduced. 
In Sec. \ref{sec:running} we show how the perturbative limit is raised in two 
3-3-1 models. In Sec. \ref{sec:deces} it is suggested a possible decoupling off all the 
scalars for energies below $\mu_{331}$. We finish in Sec. \ref{disc} with the discussion 
of our results.

\section{The running of the electroweak mixing angle}
\label{sec:sec2}

At the SU(2)$_L\otimes$U(1)$_Y$ energy scale the renormalization group equations 
dictating the running of the gauge couplings at one loop level are given by
\begin{equation}
\frac{1}{\alpha_i(\mu)}=\frac{1}{ \alpha_i(M_Z)}+\frac{1}{2\pi}
\,b_i\,\ln\left(\frac{M_Z}{\mu}\right),\;\mu\leq\mu_{331};
\label{rccgeral}
\end{equation}
with $i=1,2$ and $\alpha_2,\alpha_1$ are the coupling constants of the SU(2)$_L$, 
U(1)$_Y$ groups, respectively. The third equation involving the QCD coupling constant 
is irrelevant for our purposes here at this approximation level, and so it will not 
enter in our developments. As we have said above, for posterior use in the 
context of the 3-3-1 models we denote the gauge coupling constants 
$ \alpha_L, \alpha_X$ for the groups SU(3)$_L$, U(1)$_X$, respectively. 
In a generic gauge group the $b_i$ coefficients are given by
\begin{equation}
b_i=\frac{2}{3}\sum_{\rm  fermions}T_R(F)_i+\frac{1}{3}\sum_{\rm
scalars}T_R(S)_i-\frac{11}{3}\,C_2(G)_i
\label{bi} 
\end{equation}
for Weyl fermions and complex scalars, with the generators $T^a$ satisfying 
$\textrm{Tr}[T^a(I)T^b(I)]=T_R(I)\delta^{ab}$ where $I=F, S$. 
For $SU(N)$ $T_R(I)=1/2$ in the fundamental
representation and $C_2(G)=N$. $C_2(G)=0$ for $U(1)$. 
These numbers can be computed for other different representations with 
aid of the identity $T_R(I)(N^2-1)=C_{2R}(I)d_R(I)$, where $C_{2R}(I)$ is the 
quadratic Casimir operator and $d_R(I)$ is the representation dimension. 
For $U(1)_y$ we use $\sum T_{R1}(F,S)=\sum y^2$ where
$y=Y/2$ for the Standard Model and $y=X$ for the 3-3-1 models. 

Below $\mu_{331}$  the $\sin^2\theta_{_W}$ running is then, according 
to the Standard Model gauge group, 
\begin{equation}
\sin^2\theta_W(\mu)=\frac{1}{1+\frac{\alpha_2(\mu)}{\alpha_1(\mu)}},\;
\mu\leq\mu_{331}.
\label{sin321}
\end{equation}
We see that for a SU(2) content presenting asymptotic freedom, i. e. 
$\alpha_2\rightarrow 0$ as $\mu\rightarrow \infty$, $\sin^2\theta_{_W}$  
increases with the energy. Above the scale $\mu_{331}$ we run 
the renormalization group equations with particles in the full 
SU(3)$_L\otimes$U(1)$_N$ representations. In the minimal models when we 
consider $\mu_{331} \leq 2$ TeV, then perturbative expansion in $\alpha_X$ 
makes sense only at energy values lower than $2.4$ TeV \cite{pl331}. 
It jeopardizes the perturbative analysis of the neutral exotic vector boson $Z^\prime$ 
which is one of the model predictions, since its mass is naively expected to 
have a lower limit near this perturbative upper bound. 

We observe that for a SU(2) content $not$ presenting asymptotic freedom, i. e. with 
$\alpha_2(\mu)$ growing when $\mu\rightarrow \mu_{331}$, then $\sin^2\theta_W(\mu)$ 
decreases in the same energy interval improving the perturbative validity in such models. 
From Eqs. (\ref{rccgeral}) and (\ref{sin321}) it is deduced that the condition for 
accomplishing this is that 
\begin{equation}
b_2> \tan^2\theta_W(M_Z)b_1.
\label{condb}
\end{equation}
Inserting the $\tan^2\theta_W$ value  at the $Z$ pole, the condition between 
the coefficients turns out to be $b_2>0.3b_1$. So, if we want to satisfy it 
some additional fields in nontrivial SU(2) representations must be considered. 
In order to give a bigger contribution to $b_2$ than to $b_1$ their 
hypercharge must be small.

To be more explicit we take the 3-3-1 model with only three scalar 
triplets as our first representative of the class \cite{pt}. 
The electric charge operator ${\cal Q}$ is, 
\begin{equation}
{\cal Q} =\frac{1}{2}\left( \lambda_3-\sqrt3\lambda_8\right)+X.
\label{q}
\end{equation}
The representation content in the quark sector is 
(the numbers inside parenthesis mean their transformation properties 
under SU(3)$_C$, SU(3)$_L$ and U(1)$_X$ respectively, or under  
SU(3)$_C$, SU(2)$_L$ and U(1)$_Y$ when this is the case) 
$Q_{mL}=(d_m,\, u_m,\, j_m)^T_L\sim({\bf3}, {\bf3}^{*},- 1/3),\;m=1,2;$ 
$Q_{3L}=(u_3,\, d_3,\,J)^T_L\sim({\bf3}, {\bf 3}, 2/3)$; 
and the respective right-handed components in singlets 
$u_{\beta R}\sim({\bf3},{\bf1},2/3)$, 
$d_{\beta R} \sim({\bf3},{\bf1},-1/3) ,\,\beta=1,2,3;$ 
$J_{R}\sim({\bf3},{\bf1},5/3)$ and $j_{mR}\sim({\bf3},{\bf1},-4/3)$. 

In the scalar sector the three triplets are 
$\chi=(\chi^{-},\,\chi ^{--},\, \chi^0)^T\sim({\bf1}, {\bf 3},-1)$, 
$\rho =(\rho^{+},\,\rho ^0,\,\rho ^{++})^T\sim( {\bf1}, {\bf 3}, 1)$ and 
$\eta =(\eta^0,\,\eta _1^{-},\, \eta _2^{+})^T \sim({\bf1},{\bf3},0)$.

Finally, in this model, leptons come in triplets like 
\begin{eqnarray}
\Psi_{aL}=(\nu_a,\,l_a,\,E_a)^T_L\sim({\bf 1},{\bf 3},0),\nonumber
\label{mb}
\end{eqnarray}
where $a=e,\mu,\tau$. With their right-handed components in singlets
$l_{aR}\sim({\bf1},{\bf1},-1),\;E_{aR}\sim({\bf1},{\bf1},+1)$. 
We have omitted right-handed neutrinos since, like any neutral singlet 
representation, they are not relevant for our purposes here. The effective 
SU(2)$_L\otimes$U(1)$_Y$ theory coming from this model, when the condensation 
$\langle\chi^0\rangle=w/\sqrt2$ occurs, have the same light multiplets which 
dominate the spectrum of the Standard Model plus an extra scalar isodoublet 
$\eta =(\eta^0,\,\eta _1^{-})^T \sim({\bf1},{\bf2},-1)$. The value $w/\sqrt2$ is 
going to be identified with $\mu_{331}$.

\begin{figure}[ht] 
\begin{center} 
\leavevmode 
\mbox{\epsfig{file=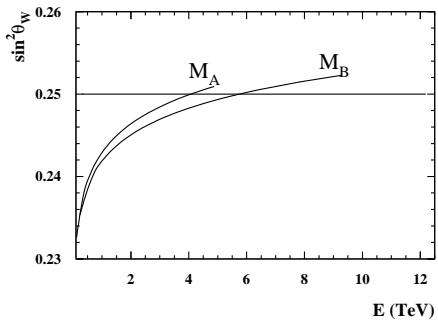,width=0.4
\textwidth,angle=0}}        
\end{center} 
\caption{Running of the electroweak mixing angle for the minimal models considering 
only light degrees of freedom under the Standard Model subgroup. Model A with 
three scalar triplets, and model B, the same as A with a scalar sextet added.}
\label{fig1} 
\end{figure}

Next we consider the renormalization group coefficients in Eq. (\ref{bi})
at the SU(2)$_L\otimes$U(1)$_Y$ level with three usual matter generations, 
taking into account at most the two nontrivial lowest dimensional SU(2) 
representations, i. e., isodoublets and isotriplets. We reserve the prefix 
$iso$ just for when we talk about SU(2)$_L\otimes$U(1)$_Y$ representations. 
Let $N_{S}$ be the number of scalar isodoublets, $N_{F}$ the number of new exotic 
fermionic isodoublets, $N_{TS}$ the number of scalar isotriplets and $N_{TF}$  
the number of new exotic fermionic isotriplets, with $Y$ being their respective 
hypercharge, then from Eq. (\ref{bi})
\begin{eqnarray}
& &b_1= \frac{1}{3}\left({Y^2_F}N_F+\frac{Y^2_S}{2}N_S+20\right)
+\frac{1}{2}\left({Y^2_{TF}}N_{TF}+\frac{Y^2_{TS}}{2}N_{TS}\right),
\label{bi1}	\\
& &b_2=\frac{1}{3}\left(N_F+\frac{1}{2}N_S+{4}N_{TF}+{2}N_{TS}-{10}\right). 
\label{bi2}	
\end{eqnarray}
When dealing with fermions in vectorial representations $N_{F}$ and $N_{TF}$ 
must be multiplied by 2. In the model above we have only two $Y^2_S=1$ 
scalar isodoublets to be considered, i. e. $N_S=2$, in the Eqs. (\ref{bi1}) 
and (\ref{bi2}) so that $(b_1,b_2)=(7,-3)$. And it is not possible to 
satisfy the condition in (\ref{condb}). In fact the $\sin^2\theta_W$ running 
shows that the value $0.25$ is reached at $4.1$ TeV as we see in Fig. (\ref{fig1}) 
(where for future reference it is also shown the same running for the model 
with a scalar sextet). 
The scale $\mu_{331}$ is then below this value and so the energy $\Lambda$ at which 
the pole is attained is such that $\Lambda\leq 4.1$ TeV. The Landau-like pole behavior 
with $\mu_{331}$ has been studied in Ref. \cite{pl331}. For example, 
if $\mu_{331}=1$ TeV we have that at energies $E\simeq 1.9$ TeV, $\alpha_{_X}>1$ and 
$M_{_{Z^\prime}}\stackrel{>}{\sim}1.9$ TeV so that the model looses its perturbative 
character below the exotic neutral vector boson $Z^\prime$ mass. 
In part this is due to the fact that the right handed components of the exotic 
quarks, $J$ and $j_{m}$, necessary to complete the SU(3)$_L$ representations have a 
value of the U(1)$_X$ charge larger than the other particles. Thus, 
they are the major contribution to the $\alpha_{_X}$ running. Under the 
assumption that they are much above the $Z^\prime$ scale, the perturbative limit is 
pushed a little above. But this brings these quarks out of experimental reach. 
We would like to pursue the possibility of keeping them observable at TeV scale. 
To make the exotic quarks very heavy means that the VEV $\langle\chi^0\rangle$ 
is very high and since this is also responsible for giving mass to all new fermions in 
the SU(3)$_L$ triplets and the new gauge bosons, 
it would bring almost all the predictions of the model out of experimental reach. 
Of course, this is not interesting from the phenomenological point of view. 

We see from Eqs. (\ref{bi1}) and (\ref{bi2}) that multiplets having large hypercharge 
take us in the opposite direction of satisfying (\ref{condb}). 
Isodoublets of vectorial leptons or/and scalars with zero hypercharge could then be 
considered. But the number of them needed would be, at least, eight of vectorial 
leptons or thirty two of scalars with components having electric charge $1/2$ 
and $-1/2$.

The simplest solution is to have exotic isotriplets. Preferentially fermionics. 
From Eqs. (\ref{bi1}) and (\ref{bi2}) one can also see that the minimal number 
of isotriplets are three, all of vectorial leptons with $Y=0$. They are written as
\begin{equation}
\vec{{\cal T}}=\left(
\begin{array}{cc}
\frac{1}{\sqrt{2}}t^0 & t^+\\
t^-& -\frac{1}{\sqrt{2}}t^{0}
\end{array}
\right),
\label{triplet}
\end{equation}
transforming as $({\bf1},{\bf3},0)$ under the Standard Model group. 
They must be color singlets, otherwise QCD asymptotic freedom would be lost. 
This isotriplet couples with the usual matter. It has interesting consequences in the 
see-saw mechanism and it does not represent cosmological troubles if they belong to 
the electroweak scale, as it was studied along with other phenomenological 
aspects in \cite{foot}. To be consistent with the $Z^0$ decay width, their components 
must be heavier than $M_{Z^0}/2$. 

Now the lowest dimensional SU(3), with U(1)$_N$ charge zero, representations which 
contain an isotriplet decomposes under the Standard Model SU(2)$\otimes$U(1)$_Y$ 
group as
\begin{eqnarray}
& &{\bf6}= {\bf3}_{-2} \oplus {\bf2}_1\oplus {\bf1}_4,\nonumber\\
& &{\bf8}= {\bf3}_0\oplus {\bf2}_{-3}\oplus {\bf2}_3\oplus {\bf1}_0.\nonumber
\label{reps}
\end{eqnarray}
The subscripts refer to their hypercharge value $Y$. The sextet gives rise to a $Y^2=4$ 
isotriplet which could also do the job but it has a $Y^2=16$ isosinglet. 
It could be assumed to be heavy not affecting the spectra below $\mu_{331}$. 
But we want to avoid such an assumption. 
It would be needed also three sextets of fermions if the $Y^2=16$ singlet components 
were heavy, plus a scalar one to generate masses. 
The octet on the other hand has the advantage of having an isosinglet and an 
isotriplet both neutrals in hypercharge which, therefore, do not contribute to $b_1$. 
But it has also two isodoublets with $Y^2=9$. They can be made heavy if a scalar 
octet is present and it condensates at the $\mu_{331}$, as we are going to discuss below.  
We shall then consider here the exotic vectorial leptons in three fermionic octets 
since they are the simplest solution. With them we also do not need to concern 
with anomalies. They have the form
\begin{equation}
{{\Xi_a}}=\left(
\begin{array}{ccc}
\frac{1}{\sqrt{2}}t^0_a+\frac{1}{\sqrt{6}}\lambda^0_a& t_a^+ & \delta^-_a\\
t_a^- & -\frac{1}{\sqrt{2}}t_a^{0}+\frac{1}{\sqrt{6}}\lambda_a^0 & \delta_a^{--}\\
\xi_a^+ & \xi_a^{++}& -\frac{2}{\sqrt{6}}\lambda_a^0 
\end{array}
\right)
\label{octetf}
\end{equation}
transforming as $({\bf1},{\bf8},0)$, where $a=1,2,3$. 

When it is introduced a representation like this, with all spinor components of the 
matter fields belonging to the same multiplet, differently from the usual 
chiral constructions, gauge invariance does allow for initial mass terms for the fields 
in these representations. Therefore, there would be a bare mass term like 
$M_{ab}{{\textrm{Tr}}}{\overline{\Xi_a}}\Xi_b$ in the Lagrangian in its symmetric fase.

To satisfy the condition in Eq. (\ref{condb}) the $Y^2=9$ lepton isodoublets 
$\delta=( \delta^-,\delta^{--})^T$ and $\xi=(\xi^{++},\xi^{+})^T$ in Eq. (\ref{octetf}) 
must then be decoupled from energies below $\mu_{331}$. This can be achieved 
with an octet representation of scalars similar to the fermions above. Calling 
this representation $\Sigma$ its form is
\begin{equation}
{{\Sigma}}=\left(
\begin{array}{ccc}
\frac{1}{\sqrt{2}}\phi^0+\frac{1}{\sqrt{6}}\varphi^0& \phi^+ & \varphi_1^-\\
\phi^- & -\frac{1}{\sqrt{2}}\phi^0+\frac{1}{\sqrt{6}}\varphi^0 & \varphi_1^{--}\\
\varphi_2^+ & \varphi_2^{++}& -\frac{2}{\sqrt{6}}\varphi^0 
\end{array}
\right)
\label{octets}
\end{equation}
transforming as $({\bf1},{\bf8},0)$. To see how the decoupling can be worked out 
let us analyze the mass spectra for these new fields.

The Yukawa interactions in the lepton sector are 
\begin{eqnarray}
-{\cal L}_Y&=&H_{ab}\epsilon_{ijk}\overline{(\Psi_{iaL})^c}\Psi_{jbL}\eta_k 
+\overline{\Psi}_{aL}(G^l_{ab}l_{bR}\rho+
G^E_{ab}E_{bR}\chi) \nonumber \\ 
&+& G^{\Xi}_{ab}\overline{\Psi}_{aL}\Xi_b\eta +  
\frac{1}{2}{\textrm{Tr}}[\overline{\Xi_a}(A_{ab}\Xi_b\Sigma + B_{ab}\Sigma\Xi_b 
+ M_{ab}\Xi_b)] 
\nonumber \\ 
&+& H.c. ,
\label{yu1} 
\end{eqnarray}
where repeated indices mean summation. $H_{ab}$ is an antisymmetric matrix with  
$G^l_{ab}$, $G^E_{ab}$, $G^{\Xi}_{ab}$ general complex matrices and 
$A_{ab}$, $B_{ab}$, $M_{ab}$ real symmetric matrices which will be assumed 
diagonal for simplicity. 

Then $Y^2=9$ isodoublets in $\Xi_a$ decouple when $\langle\Sigma\rangle=
\frac{1}{\sqrt{6}} \langle\varphi^0\rangle diag( 1, 1, -2)$, when the following 
relations are satisfied
\begin{eqnarray}
& & (A_{ab}+B_{ab}) \frac{\langle\varphi^0\rangle}{\sqrt{6}}+M_{ab}\simeq m_{_{SM}},
\nonumber \\ \nonumber \\ 
& & m_{_{SM}}-3A_{ab}\frac{\langle\varphi^0\rangle}{\sqrt{6}}\simeq\mu_{331},
\nonumber\\ \nonumber \\ 
& & m_{_{SM}}-3B_{ab}\frac{\langle\varphi^0\rangle}{\sqrt{6}}\simeq\mu_{331}.
\label{vinc} 
\end{eqnarray}
Where $m_{_{SM}}$ is a mass scale belonging to the Standard Model mass spectrum.
The first relation above leaves the isotriplets $\vec{{\cal T}}_a$ in ${\Xi}_a$ 
with appropriate masses below $\mu_{331}$. The last two relations in (\ref{vinc}) 
ensure that the masses of the components of the isodoublets $\delta$ and $\xi$ 
are heavy enough, since $\langle\Sigma\rangle$ breaks the electroweak SU(3)$_L$ 
symmetry down to the SU(2)$_L$ group $\langle\varphi^0\rangle\simeq\mu_{331}$ and 
the isodoublets are, in this way, decoupled of the Standard Model particle 
mass spectrum. 
The isosinglets $\lambda^0_a$ in the octet $\Xi_a$ have then masses of order 
$m_{\lambda^0_a}\simeq M_{aa}-m_{_{SM}}$ which is the same order of $\mu_{331}$ 
according to the Eqs. (\ref{vinc}). 
The simple analysis presented here is not complete but it is sufficient for the 
purposes of showing that the higher hypercharge fields can be heavy. Thus, only the 
vector lepton isotriplets $\vec{{\cal T}}_a$ together with the known fields 
contribute to the renormalization group equations at energies $E<\mu_{331}$.   

In the last stage of symmetry breakdown when the Standard Model group reduces 
to the electromagnetic abelian gauge group, U(1)$_{em}$, we see from the term 
$G^{\Xi}_{ab}\overline{\Psi}_{aL}\Xi_b\eta$ in Eq. (\ref{yu1}) that there will be 
a mixing between the know and the exotic leptons. It could be avoided by 
imposing some sort of discrete symmetry. 
Their phenomenological consequences will be carried out elsewhere.

Finally, we should say that condensation of $\phi^0$ in $\Sigma$ is very 
constrained by the $\rho$ parameter relating the $W^{\pm}$ and $Z$ masses. 

We are ready now to show how the perturbative limit is improved with these 
representations.

\section{The perturbative limit}
\label{sec:running}

We now turn out to look how the perturbative limit is characterized by 
the scale $\mu_{331}$ breaking the SU(3)$_L\otimes$U(1)$_X$ symmetry. The relevant 
running equation is that of the coupling constant $\alpha_X$ coming from the 
abelian factor U(1)$_X$. This running is 
\begin{equation}
\frac{1}{\alpha_X(\mu)}=\frac{1}{\alpha_X(\mu_{331})}+
\frac{1}{2\pi}b_X\,\ln\left(\frac{\mu_{331}}{\mu}\right),\;\mu\geq\mu_{331}.
\label{alphaxrun}
\end{equation}
$b_X$ is computed considering the fields in the full 3-3-1 representation. 
Eq. (\ref{alphaxrun}) starts with the value $\alpha_X(\mu_{331})$ which can be determined 
using Eqs. (\ref{polo}) and (\ref{sin321}) and the fact that 
$\alpha_2(\mu_{331})=\alpha_L(\mu_{331})$. That is, 
\begin{equation}
\frac{1}{\alpha_X(\mu_{331})}=\frac{1}{\alpha_1(\mu_{331})}-\frac{3}{\alpha_2(\mu_{331})}.
\label{alphaxmu331}
\end{equation}
Perturbative expansion does not make sense any more when a scale $\mu=M^\prime$ such 
that $\alpha_X(M^\prime)=1$ is reached. We have then
\begin{equation}
M^\prime=\mu_{_{331}} \left[\frac{M_{_Z}}{\mu_{_{331}}} \right]^{\frac{b_1-3b_2}{b_{_X}}}
e^{\frac{2\pi}{b_{_X}}\left[\frac{1}{\alpha(M_{_Z})}(1-4 \sin^2\theta_W(M_{_Z}))-1
\right]},
\label{pololandau}
\end{equation}
$\alpha(M_{_Z})=1/128$ is the fine structure constant at the $Z$ pole, with 
$M_Z=91.2$ GeV. We can now see how the perturbative limit is changed with 
$\mu_{_{331}}$. 

\begin{figure}[ht] 
\begin{center} 
\leavevmode 
\mbox{\epsfig{file=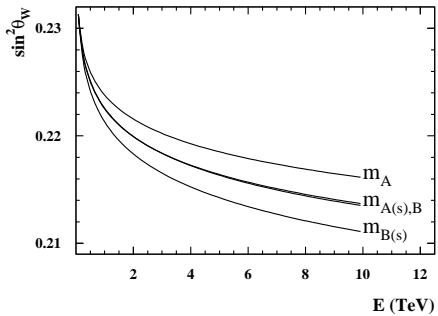,width=0.4\textwidth,angle=0}}        
\end{center} 
\caption{Running of the electroweak mixing angle for the extended models for energies 
E$\leq\mu_{_{331}}$. Model A, 
$m_A$, with only three scalar triplets, and model B, $m_B$, with a scalar sextet. 
(s) means that the $Y^2=9$ scalars are disregarded. The models $m_A(s)$ and $m_B$ 
have degenerate curves.}
\label{fig2} 
\end{figure}

The additional representations to be considered are then the three leptonic 
SU(3)$_L$ octets $\Xi_a$, ($a=1,2,3$), like (\ref{octetf}) which leave 
three leptonic isotriplets like (\ref{triplet}) at energies $E\leq \mu_{331}$, 
i. e. $N_{TF}=3$ in Eqs. (\ref{bi1}) and (\ref{bi2}). 
The scalars in $\Sigma$ can be assumed 
to contribute with a $Y^2=0$ isotriplet and two $Y^2=9$ isodoublets. 
Thus from Eqs. (\ref{bi1}) and (\ref{bi2}) we have $(b_1,b_2)= (10,6)$, 
which obviously satisfies the condition (\ref{condb}). In Fig. (\ref{fig2}) 
the $\sin^2\theta_W$ running in this model is shown(again for future 
reference it is also shown the curves in the model with a scalar sextet). 

From Eq. (\ref{bi}) we also have 
$b_X=26$. For example, if $\mu_{_{331}}=1$ TeV then $M^\prime=16.8$ TeV; if 
$\mu_{_{331}}=3$ TeV then $M^\prime=70.6$ TeV. Disregarding the $Y^2=9$ scalar 
isodoublets $(b_1,b_2)= (7,17/3)$ and if $\mu_{_{331}}=1$ TeV then $M^\prime=20.2$ TeV; 
if $\mu_{_{331}}=3$ TeV then $M^\prime=92.3$ TeV. 
The perturbative limit is still improved in this case because $b_1$ is lowered. 
We see then that the model is made perturbatively much more viable with the 
addition of octet representations.

\subsection{The model with a scalar sextet}

\label{subsec:mb}

Another representation content in the leptonic sector is possible in accordance with 
the electric charge operator Eq. (\ref{q}). The positron and his cousins occupying the 
third component of the leptonic triplets which are represented by \cite{331}, 
\begin{eqnarray}
\Psi_{aL}=(\nu_a,\,l_a,\,l^c_a)^T_L\sim({\bf1},{\bf3},0) \nonumber
\label{tlmb}
\end{eqnarray}
with $a=e,\mu,\tau$. A scalar sextet 
\begin{eqnarray} 
S=\left(\begin{array}{lll} 
\sigma^0_1 & h^-_1 & h^+_2\\ 
h^-_1 & H^{--}_1 & \sigma^0_2 \\ 
h^+_2 & \sigma^0_2 & H^{++}_2  
\end{array}\right)\nonumber
\label{sextet} 
\end{eqnarray} 
transforming as $({\bf1},{\bf6},0)$ is necessary in the minimal representation 
content to generate all the lepton masses when $\langle\sigma^0_2\rangle\not=0$. 
$\langle\sigma^0_1\rangle\not=0$ could give a Majorana mass to the neutrinos. 

The Yukawa interactions in the lepton sector are now
\begin{eqnarray}
-{\cal L}^s_Y&=&H_{ab}\epsilon_{ijk}\overline{(\Psi_{iaL})^c}\Psi_{jbL}\eta_k 
+G^{\Psi}_{ab}\overline{\Psi}_{aL}S(\Psi_{ibL})^c \nonumber \\ 
&+& G^{\Xi}_{ab}\overline{\Psi}_{aL}\Xi_b\eta +  
F_{ab}\epsilon_{ijk}\overline{\Psi_{iaL}}(S^{\dagger}\Xi_b)_{jk}\nonumber \\ 
&+&\frac{1}{2}{\textrm{Tr}}[\overline{\Xi_a}(A_{ab}\Xi_b\Sigma + B_{ab}\Sigma\Xi_b 
+ M_{ab}\Xi_b)] \nonumber \\ 
&+&H.c.
\label{yu2} 
\end{eqnarray}
And the same decoupling of the $Y^2=9$ lepton doublets can be worked out as before. 
Mixing between the know and exotic leptons happens through 
$G^{\Xi}_{ab}\overline{\Psi}_{aL}\Xi_b\eta $ and 
$F_{ab}\epsilon_{ijk}\overline{\Psi_{iaL}}(S^{\dagger}\Xi_b)_{jk}$ in this case. 
As before, it could be avoided by imposing some sort of discrete symmetry. 

This model has the following additional multiplets 
under the SU(3)$_C\otimes$SU(2)$_L \otimes $U(1)$_Y$ symmetry besides 
those in the three triplets model: 
one isodoublet $(h_2^+,\,\sigma_2^0)\sim({\bf1},{\bf2},+1)$, a nonhermitian isotriplet
$(H_1^{--},h_1^-,\sigma_1^0)\sim({\bf1},{\bf3},-2)$, and an isosinglet
$H_2^{++}\sim({\bf1},{\bf1},+2)$. Then, at energies below $\mu_{_{331}}$ three scalar 
isodoublets with $Y^2=1$ and two scalar isodoublets with $Y^2=9$ are taken into 
account but now we have one more scalar isotriplet plus that coming from the scalar 
octet. The other scalar fields including the isosinglet $H_2^{++}$ are taken into 
account only at energies above $\mu_{_{331}}$. In this case the inclusion of the 
isosinglet $H_2^{++}$ does not affect so significantly the running below 
$\mu_{_{331}}$. Therefore, $N_S=3$ with $Y^2=1$, $N_S=2$ with $Y^2=9$, $N_{TS}=1$ 
with $Y^2=4$, $N_{TS}=1$ with $Y^2=0$ and $N_{TF}=3$ with $Y^2=0$.
So that from Eqs. (\ref{bi1}) and (\ref{bi2}) we have $(b_1,b_2)= (67/6,41/6)$.
Obviously, this satisfies the condition (\ref{condb}). From Eq. (\ref{bi}) we also have 
$b_X=22$ since there is no right-handed charged leptons to be counted. 
For example, if $\mu_{_{331}}=1$ TeV then $M^\prime=32.4$ TeV; if $\mu_{_{331}}=3$ TeV
then $M^\prime=155.0$ TeV. Disregarding the $Y^2=9$ scalar isodoublets coming from 
the octet $\Sigma$, $(b_1,b_2)= (49/6,13/2)$, so if $\mu_{_{331}}=1$ TeV then 
$M^\prime=40.3$ TeV; if $\mu_{_{331}}=3$ TeV then $M^\prime=213.0$ TeV. Again, 
perturbative limit is still improved in this case since $b_1$ is lowered.

\section{Possible decoupling of all scalars at energies $E<\mu_{_{331}}$}
\label{sec:deces}

In the preceding discussions it was considered that some scalar multiplets remain 
light enough, i. e., with masses much below the $\mu_{331}$ scale, so that they could 
participate in the counting for the coefficients in Eqs. (\ref{bi1}) and (\ref{bi2}). 
It only happens if there is a sort of fine tuning among the scalar self-interaction 
couplings. The reason for this comes from the fact that the scalar octet has 
interactions with all the scalar multiplets. The scalar potential for the three triplets 
model plus the octet is then
\begin{eqnarray}
V & = & V_{T}+ m_{3}\eta^\dagger\Sigma \eta+ 
m_{4}\rho^\dagger\Sigma \rho+
m_{5}\chi^\dagger\Sigma \chi\nonumber\\
&+& {\textrm{Tr}}[\Sigma^2(\lambda_{1}\Sigma^2+ m_{1}\Sigma-m_{2}^2)]
+\lambda_{2}({\textrm{Tr}}[\Sigma^2])^2
\nonumber\\
&+& { \textrm{Tr}}[\Sigma^2]\left(
\lambda_{3} \eta^\dagger \eta+ 
\lambda_{4} \rho^\dagger \rho+
\lambda_{5} \chi^\dagger \chi\right).
\label{pe1}
\end{eqnarray}
\noindent 
Where $V_{T}$ stands for the collection of all terms  involving only $\eta$, $\rho$ 
and $\chi$. Note that we cannot impose invariance under $\Sigma\rightarrow-\Sigma$, 
since it would also forbid the interaction between $\Sigma$ and the $\Xi_a$'s 
preventing the mechanism of decoupling the lepton isodoublets $\delta$ and $\xi$ of 
the Standard Model scale. The last term in this potential gives a  quadratic mass 
term of order $\langle\varphi^0\rangle^2$ for the triplets unless $\lambda_{3}$, 
$\lambda_{4}$ and $\lambda_{5}$ are fine tuned. Thus, it can happen 
that no single scalar be part of the spectrum below  $\mu_{331}$. 
In fact, the only massive scalar state which could be free of mass contributions 
coming from $\langle\Sigma\rangle$ would belong to the neutral scalar imaginary part, 
the pseudoscalar. 
But in the minimal model, i. e., without regarding the octets, there is just one 
massive pseudoscalar and it might be heavy, since its mass is of order 
$\langle\chi^0\rangle$. Of course, careful mass spectral analysis should be done 
but the arguments we just have given indicate that it really should happen. 
Because of the reason that the major contributions to the renormalization 
group equations are due to the fermions, it is not expected that the perturbative 
limit will change significantly when the scalars are omitted below $\mu_{331}$. 
But to produce any fundamental scalar would require energies $E\geq\mu_{331}$. 

\section{Discussion}
\label{disc}

We have shown that the addition of octet representations can turn 3-3-1 models perturbatively 
viable at the upcoming TeV energy scale. The main effect is that they predict 
a running to the electroweak angle, or better $\sin^2\theta_W$, such that it decreases 
with the increasing of the energy until the scale $\mu_{331}$ which characterizes the 
appearance of a SU(3)$_L\otimes$U(1)$_X$ electroweak symmetry. It could be tested providing a 
direct test of exclusion for the extended models we have presented here. With the electroweak 
angle evolving in this way the U(1)$_X$ coupling constant, $\alpha_X$, starts to run 
with a value which is lower than the one appearing in the minimal models. From this point 
$\sin^2\theta_W$ begins to increase with energy but it will reach values near $0.25$ 
at energies bigger than tens of TeV.

The same behavior could be accomplished with other representations different from  
octets. But these representations would be in general bigger than the ones we have 
considered here, and they would need more complicated mechanisms to decouple isomultiplets 
with large hypercharge. If instead we consider exotic leptons in sextets it is 
necessary to assume that the isosinglets with $Y^2=4$ belonging to them are decoupled 
at energies $E\leq\mu_{331}$. Thus octets are simpler and more convenient. 

Although the additional representations turn the SU(3)$_L$ gauge coupling, 
$\alpha_{3L}$, looses its asymptotic freedom, it becomes bigger than one at 
energies much higher than $M^\prime$. Thus, it does not represent a problem.

Symmetry breakdown at the scale $\mu_{331}$ is marked with the appearance of four new 
charged vector bosons, two of them carrying two units of electric charge, and the 
exotic neutral one $Z^\prime$. Different from the exotic charged vector bosons, 
this neutral one does not receive contribution to its mass due to the scalar octet 
and it is bounded according to 
\begin{eqnarray}
M_{_{Z^\prime}} \stackrel{>}{\sim}2\left[\alpha_{_2}(\mu_{_{331}})
\frac{1-\sin^2\theta_W(\mu_{_{331}})}{1-4\sin^2\theta_W(\mu_{_{331}})}
\right]^{\frac{1}{2}} \mu_{_{331}}.
\label{zlmass}
\end{eqnarray}
This is valid for both models here. 
It is assumed that $\sqrt2\langle\chi_0\rangle\simeq\mu_{_{331}}$, which is quite 
reasonable since as $\langle\Sigma\rangle$, $\langle\chi\rangle$ also breaks the 
symmetry down to the Standard Model group. $M_{_{Z^\prime}}$ grows then linearly 
with $\mu_{_{331}}$ so that its production will mark, according to these models, 
the scale at which SU(3)$_W$ should be taken into account. Thus, the $Z^\prime$ 
boson would have a mass which could be tens of TeV or even much less,  
if other phenomenological constraints allow for it.

Finally, we should say that supersymmetry does not relieve the perturbative 
limit in the minimal model versions \cite{pl331}. It brings more degrees of freedom 
with the supersymmetric partners and so it worsens the problem. But for the extended 
models presented here it is expected that it makes the value of $M^\prime$ in 
Eq. (\ref{pololandau}) increase.

The author thanks C. A de S. Pires, P. S. R. da Silva and V. Pleitez for their 
useful comments and M. Caballero for reading the manuscript. 
This work has been supported by  FAPESP under the process 01/13607-3.


\begin{thebibliography}{99}

\bibitem{331} F. Pisano and V. Pleitez, Phys. Rev. D {\bf46} (1992) 410;
R. Foot, O. F. Hernandez, F. Pisano and V. Pleitez, Phys. Rev. D {\bf 47} 
(1993) 4158;
P. H. Frampton, Phys.\ Rev.\ Lett.\ {\bf 69} (1992) 2889.
\bibitem{rasin} P. H. Frampton and A. Rasin, Phys. Lett. {\bf B482}, 129 
(2000). 
\bibitem{dion} B. Dion, T. Gregoire, D. London, L. Marleau and H. Nadeau, 
Phys. Rev. D {\bf 59}, 075006 (1999).
\bibitem{maisbi} M. B. Tully and G. C. Joshi, Phys. Lett. {\bf B466}, 333 (1999); 
E. M. Gregores, A. Gusso and S. F. Novaes, Phys. Rev. D {\bf64}, 015004 (2001); 
J. L. Garcia-Luna, G. Tavares-Velasco, J. J. Toscano, Phys. Rev. D {\bf69}, 093005 (2004).
\bibitem{pt} V. Pleitez and M. D. Tonasse, Phys. Rev. D {\bf48}, 2353
(1993). 
\bibitem{pdg} K. Hagiwara {\it et el.}, (Particle Data Group), Phys. Rev. D
{\bf66}, 010001 (2002). 
\bibitem{phf98} P. H. Frampton, Int. J. Mod. Phys. {\bf A13}, 2345
(1998), P. H. Frampton, Mod. Phys. Lett. {\bf A18}, 1377 (2003).
\bibitem{pl331} A. G. Dias, R. Martinez, V. Pleitez, hep-ph/0407141. 
\bibitem{outros331} J. C. Montero, F. Pisano and V. Pleitez, Phys. Rev. D {\bf47}, 
2918 (1993);  
R. Foot, H. N. Long and T. A. Tran,  Phys. Rev. D {\bf50}, R34 (1994);  
H. N. Long,  Phys. Rev. D {\bf53}, 437 (1996);  
\bibitem{outros} A. G. Dias, V. Pleitez, and M. D. Tonasse, Phys. Rev. D
{\bf67}, 095008 (2003); A. G. Dias and V. Pleitez, Phys. Rev. D {\bf69}, 077702 (2004); 
H. N. Long, N. Q. Lan, Europhys. Lett. {\bf68}, 571 (2003);
A. G. Dias, C. A. de S. Pires, and P. S. Rodrigues da Silva, 
Phys. Rev. D {\bf68}, 115009 (2003); 
G. Tavares-Velasco and J. J. Toscano, Phys. Rev. D {\bf65}, 013005 (2002); 
C. A. de S. Pires and P. S. Rodrigues da Silva, Eur. Phys. J. C{\bf36}, 397 (2004); 
M. A. Perez, G. Tavares-Velasco and J. J. Toscano, Phys. Rev. D {\bf69}, 115004 (2004); 
J. C. Montero, V. Pleitez and M. C. Rodriguez, Phys. Rev. D {\bf70}, 075004 (2004); 
G. Tavares-Velasco and J. J. Toscano,  Phys. Rev. D {\bf70}, 053006 (2004); 
J. A. Rodriguez and M. Sher, hep-ph/0407248. 
\bibitem{foot} R. Foot, H. Lew, X.-G. He, G. C. Joshi, Z. Phys. C {\bf44}, 
441 (1989)
\bibitem{seesaw331} J. C. Montero, C. A. de S. Pires and V. Pleitez, 
Phys. Rev. D {\bf65}, 095001 (2002). The electric charge atribution of the lepton 
octet must be corrected in this work. But it does not affect their subsequent
conclusions.
\end{thebibliography}
\end{document}